\documentclass[floatfix,intlimits,pra,showkeys,showpacs,twocolumn]{revtex4-1}

\usepackage{amsfonts,amsmath,amssymb}
\usepackage{dcolumn}
\usepackage[dvips]{graphicx}
\PassOptionsToPackage{hyphens}{url}
\usepackage{hyperref}

\begin{document}

\title{Design and construction of a multistage Zeeman decelerator for crossed molecular beams scattering experiments}

\author{Theo Cremers}
\author{Niek Janssen}
\author{Edwin Sweers}
\author{Sebastiaan Y.T. van de Meerakker}
\email{basvdm@science.ru.nl}

\affiliation{Radboud University, Institute for Molecules and Materials, Heijendaalseweg 135, 6525 AJ Nijmegen, the Netherlands}

\date{\today}

\begin{abstract}
Zeeman deceleration is a relatively new technique used to obtain full control over the velocity of paramagnetic atoms or molecules in a molecular beam. We present a detailed description of a multistage Zeeman decelerator that has recently become operational in our laboratory [Cremers \emph{et al.}, Phys. Rev. A 98, 033406 (2018)], and that is specifically optimized for crossed molecular beams scattering experiments. The decelerator consists of an alternating array of 100 solenoids and 100 permanent hexapoles to guide or decelerate beams of paramagnetic atoms or molecules. The Zeeman decelerator features a modular design that is mechanically easy to extend to arbitrary length, and allows for solenoid and hexapole elements that are convenient to replace. The solenoids and associated electronics are efficiently water cooled and allow the Zeeman decelerator to operate at repetition rates exceeding 10~Hz. We characterize the performance of the decelerator using various beams of metastable rare gas atoms. Imaging of the atoms that exit the Zeeman decelerator reveals the transverse focusing properties of the hexapole array in the Zeeman decelerator.
\end{abstract}

\pacs{37.10.Mn, 37.20.+j}
\maketitle

\section{Introduction}\label{sec:intro}

The crossed molecular beams technique is ideally suited to study fundamental interactions that take place when molecules and atoms collide \cite{Lee:Science236:793}. In order to validate the latest theoretical models and approaches, experiments with the highest level of precision are required. Ideally, the collision partners initially occupy only one quantum state, and they collide with a continuously tunable collision energy and high inherent energy resolution. To generate molecular collision partners, seeded supersonic beams are often used as a source, as they provide samples of molecules with relatively narrow velocity distributions and low rotational and vibrational temperatures \cite{Depaul:JPC97:2167}. The collision energy can be changed in discrete steps by using different seed gases, or varied continuously over a wide range by changing the angle between the two beams \cite{Chefdeville:Science341:06092013, Henson:Science338:234, Osterwalder:EPJ-TI2:10}.

However, even when the best molecular beam sources are used, the energy resolution in a collision experiment is mainly limited by the velocity spread of the parent beams. Additional control over molecular beams can be obtained by manipulating them with magnetic or electric fields \cite{Stuhl:ARPC65:501,Brouard:CSR43:7279,Meerakker:CR112:4828}. Neutral particles with an electric or magnetic dipole moment experience a force when they encounter inhomogeneous electric or magnetic fields, through the Stark and Zeeman effect, respectively. These forces can be used to focus or deflect molecules, but can also be used to change the forward velocity and velocity distribution of a packet of molecules. The latter was pioneered in the late 90's, and resulted in the construction of the so-called Stark decelerator, that consists of an array of high-voltage electrodes to manipulate beams of neutral polar molecules \cite{Bethlem:PRL83:1558}. With the Stark decelerator, packets of molecules can be produced with a computer-controlled velocity, a narrow velocity spread, and almost perfect quantum-state purity. Almost a decade later, the first multistage Zeeman decelerators were developed for particles with a magnetic dipole moment \cite{Vanhaecke:PRA75:031402,Narevicius:NJP9:358}.

The Stark deceleration technique was first used in a crossed beams scattering experiment in 2006 \cite{Gilijamse:Science313:1617}, and has been used in a variety of collision experiments ever since \cite{Zastrow:NatChem6:216, Onvlee:CPC17:3583, Vogels:SCIENCE350:787, Vogels:NatureChem10:435, Gao:NatureChem10:469}. The Zeeman deceleration technique has thus far successfully been used in novel trapping \cite{Akerman:PRL119.073204, Liu:PRL118:093201} and spectroscopic experiments \cite{Jansen:JMolSpec322:9}, but its application in molecular scattering experiments is still in its infancy. Recently, a magnetic synchrotron was proposed that allows for the storage of multiple packets of hydrogen atoms for merged beam collisions with a variety of non-magnetic species \cite{Poel:NJP17:055012}.

Here we give a detailed description of a multistage Zeeman decelerator that is specifically optimized for crossed molecular beams scattering experiments. The 3-meter long decelerator has become operational in our laboratory recently \cite{Cremers:PRA98:033406} and consists of an alternating array of 100 solenoids and 100 hexapoles, such that the transverse and longitudinal motions of particles traveling through the decelerator are decoupled and can be controlled separately. The solenoids are made of copper capillary and are internally cooled using a flow of liquid, in principle affording repetition rates of the experiment exceeding 10 Hz. Electrical currents up to 4.5 kA are pulsed through the solenoids using cost-effective components based on field-effect transistors (FETs), whereas the hexapoles consist of an array of permanent magnets. The decelerator features a modular design, and can easily be extended or shortened to arbitrary length by adding or removing modules. We present a detailed description of the decelerator design and construction, focusing primarily on the mechanical and electronical aspects. We characterize the focusing properties of the decelerator using beams of metastable He, Ne, Ar or Kr atoms, that are guided through the decelerator and imaged on a screen upon exiting the decelerator.

This paper is organized as follows: In Section \ref{sec:prin} we explain the operation principle of the multistage Zeeman decelerator presented here. Section \ref{sec:mech} describes the mechanical design of the apparatus in detail, while Section \ref{sec:elec} provides a detailed account on the electronics that power the solenoids. Section \ref{sec:trans} provides a comparison between the solenoid and hexapole contribution to the transverse forces in the decelerator. In Section \ref{sec:setup} the experimental setup that we use to characterize the decelerator is explained. The results of these experiments are presented in Section \ref{sec:res}. Finally, in Section \ref{sec:sum} the paper is summarized and the prospects for our Zeeman decelerator in scattering experiments is discussed.

\section{Experimental}

\subsection{Operation principle}\label{sec:prin}
We first give a brief description of the operation principles of the decelerator; a full description can be found elsewhere \cite{Cremers:PRA95:043415}. In the following, we consider paramagnetic particles that are in a low-field-seeking (LFS) quantum state, i.e., particles whose Zeeman energy increases when moving into a stronger magnetic field. This increase in potential energy is compensated by a loss in kinetic energy. Upon leaving the field the particles regain an equal amount of kinetic energy. However, the amount of deceleration and acceleration can be controlled by turning on or off the magnetic field at specific times.

\begin{figure}[!tb]
\centering
\resizebox{1.0\linewidth}{!}
{\includegraphics{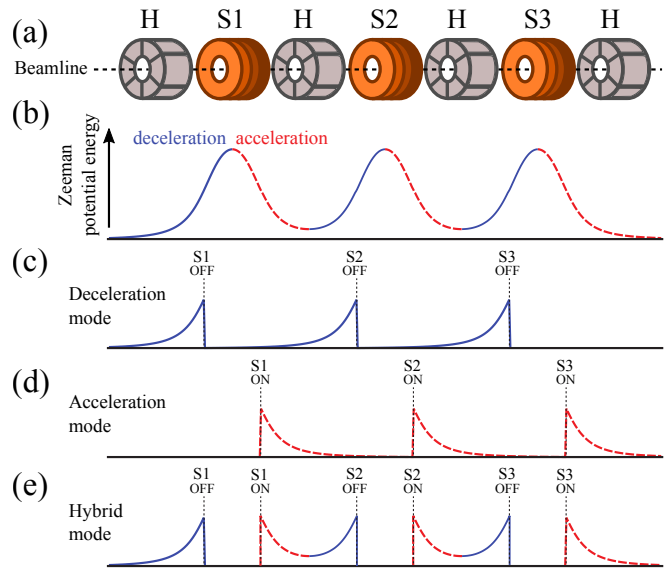}}
\caption{(\emph{a}) Schematic representation of the array of alternating hexapoles (H) and solenoids (S). (\emph{b}) Zeeman potential energy experienced by a low-field-seeking atom or molecule traveling along the beam axis for activated solenoids. The decelerator can be operated in a deceleration mode (\emph{c}), an acceleration mode (\emph{d}), and a hybrid mode (\emph{e}). For each mode, an example of the Zeeman energy experienced by the synchronous particle is illustrated. Blue and red curves indicate the part of the potentials that decelerate and accelerate low-field-seeking particles, respectively.}
\label{fig:modexplain}
\end{figure}

In order to better understand the principle of longitudinal velocity control using pulsed solenoids, it is instructive to consider a so-called synchronous particle that can be though of as the center of the packet of molecules manipulated by the decelerator \cite{Wiederkehr:PRA82:043428,Cremers:PRA95:043415,Dulitz:PRA91:013409}. This synchronous particle is only described in one dimension, along the beamline, as illustrated in Fig. \ref{fig:modexplain}(a). When all the solenoids are active, the Zeeman potential energy of this hypothetical particle as a function of longitudinal position is shown in Fig. \ref{fig:modexplain}(b). By turning solenoids on or off each time the synchronous particle reaches the same point relative to the solenoid center, the kinetic energy of this particle will change an equal amount per stage, assuming instantaneous switching of the solenoid fields. In reality, the solenoids have significant rise and fall times of the currents, and the exact amount of deceleration per solenoid depends on the velocity of the synchronous particle \cite{Dulitz:PRA91:013409}. The kinetic energy of the particle can be decreased or increased with the so-called 'deceleration mode' and 'acceleration mode', respectively. Fig. \ref{fig:modexplain}(c) and (d) show examples of the potential energy of the synchronous particle during deceleration and acceleration mode. The vertical dashed lines mark points where the solenoids are switched on or off, as labeled.

In addition to controlling the longitudinal velocity of the synchronous particle, switching the solenoids off during an upwards slope, or on during a downward slope, will result in longitudinal focusing of the nearby particles in the molecular beam \cite{Wiederkehr:PRA82:043428,Cremers:PRA95:043415,Dulitz:PRA91:013409}. Relative to the synchronous particle, non-synchronous particles are either accelerated if they lag behind or decelerated if they are ahead. This focusing effect is limited by the strength of the magnetic field, and the speed at which the current rises or falls. The position of the synchronous particle at the time of switching also has an influence; the closer to the apex of the potential hill, the weaker the focusing effect.

The two modes can also be combined into a so-called 'hybrid mode' (as depicted in Fig. \ref{fig:modexplain}(e)), which maximizes longitudinal focusing of the molecules. This mode can either be used to make small changes in longitudinal mean velocity, or to guide the packet of molecules at constant mean velocity. The acceleration, deceleration and hybrid modes of operation together allow for the production of a well-defined packet of atoms or molecules within a large range of final mean velocities.

\subsection{Mechanical implementation}\label{sec:mech}

\begin{figure}[!tb]
\centering
\resizebox{1.0\linewidth}{!}
{\includegraphics{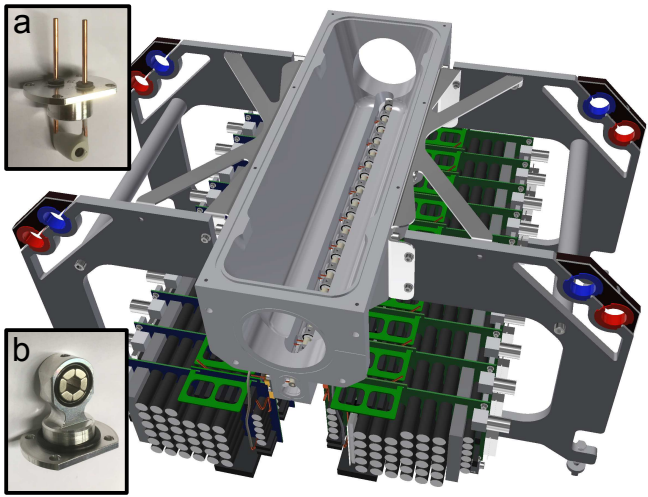}}
\caption{Projection of the 3-D model of one multistage Zeeman decelerator module, including the supporting frame and the printed-circuit boards. The cooling tubes and the lid are not included for clarity. The module contains 20 solenoids that are inserted from the sides and 19 hexapoles that are connected from below. The photographs (a) and (b) show respectively a solenoid and a hexapole inside an aluminium flange.}
\label{fig:Zeemanview}
\end{figure}

\begin{figure*}[!htb]
\centering
\resizebox{1.0\linewidth}{!}
{\includegraphics{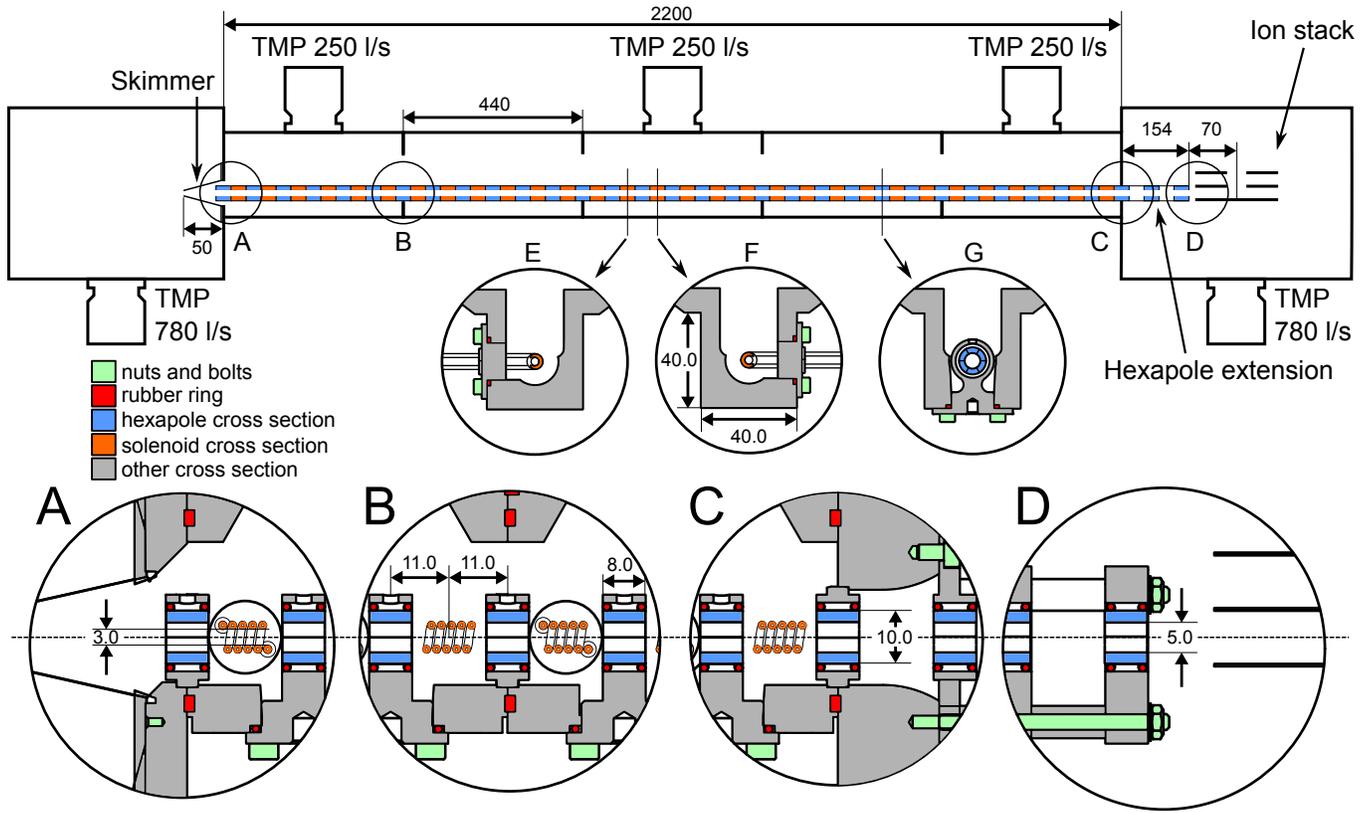}}
\caption{Schematic representation of the multistage Zeeman decelerator. Distances are given in mm. The top image shows the Zeeman decelerator together with the source and detection chamber. In this image, four areas of interest are depicted in detailed longitudinal cross sections. These areas are: (A) Connection between source chamber and decelerator module. (B) Connection between two modules. (C) Connection between module and detection chamber. (D) End of the hexapole extension and part of the ion stack. Three transverse cross sections below the main diagram show solenoids mounted from the left (E) and right (F), and a hexapole mounted from below (G).}
\label{fig:Zeemancross}
\end{figure*}

The multistage Zeeman decelerator presented here features a modular design that allows for an easy extension to arbitrary length. Each module houses 20 solenoids and 19 hexapoles. In Fig. \ref{fig:Zeemanview} a depiction of one such module is shown. For easy maintenance and replacement of parts, the solenoids and hexapoles are mounted individually using an aluminium mounting flange, as can be seen in Fig. \ref{fig:Zeemanview}(a) and (b). The decelerator consists of five modules, yielding a total of 100 deceleration stages. In Fig. \ref{fig:Zeemancross} a schematic drawing of the complete decelerator is shown, together with longitudinal cross sections detailing the connections between the source chamber, the decelerator modules and the detection chamber, as well as transverse cross sections that illustrate the positioning of the hexapoles and solenoids. A number of additional hexapoles are installed in the region between the end of the decelerator and the detection region, in order to maintain the transverse particle distribution of the decelerated beam into the detection chamber, as will be further described in section \ref{sec:setup}.

A single module is made from a 150 mm x 150 mm x 440 mm block of aluminium. First, it is modified into roughly the shape of the chamber by a computer controlled milling machine. The final alignment of the decelerator is purely mechanical, therefore the precision of the solenoid and hexapole ports, as well as the connection between adjacent modules is critical. This high level of precision can be achieved by milling all critical sections in a single session. For this, the aluminium block is fixed in a 5-axis computer controlled milling machine, such that the position of each face and threaded hole is well-defined to the accuracy of the 5-axis machine.

Adjacent modules use a stainless steel ring as a dowel, fixing two modules together and aligning them (see Figure \ref{fig:Zeemancross}B). This ring holds a hexapole, in order to present an uninterrupted sequence of solenoids and hexapoles throughout the decelerator. Similar dowels are used at the entrance and exit of the decelerator, where two custom CF200 flanges interface the decelerator to a commercial source and detection vacuum chamber, respectively. For ease of installation, the aluminium modules are connected to a set of braces that are designed to rest on two girders that run parallel with the decelerator. These braces also hold the cooling liquid manifolds running alongside the decelerator.

The solenoids and hexapoles have a center-to-center distance of 11 mm, limited by the combined length of a single solenoid and hexapole, with approximately a 1-mm spacing between them. The hexapoles are inserted into the module from the bottom, whereas the solenoids are inserted from either sides of the module in an alternating manner to facilitate sufficient space for the installation of a printed-circuit board (PCB) for each solenoid (see also the transverse cross sections of a module shown in Fig. \ref{fig:Zeemancross}(E-G)). The top of a module is closed with a lid that has two CF63 and one CF100 flange ports, which can be used to install turbo-molecular pumps, pressure gauges, or auxiliary equipment. In our setup, only three of the five modules are equipped with a 250 l/s pump, which yields sufficient pumping capacity to run the experiment at a 10-Hz repetition rate.

The hexapoles are made from six arc-shaped permanent magnets that form a ring with alternating radial magnetization. The commercially available magnets (China Magnets Source Material LTD), consist of grade N30M sintered neodymium. Magnets of this grade typical have a surface magnetic field of about 1.1 T. The magnet ring has an outer diameter of 10 mm, an inner diameter of 5 mm and has a length of 8 mm. The magnet assemblies are fixed in place with two rubber rings in an aluminium mounting flange (see Fig. \ref{fig:Zeemanview}(b)). A disadvantage of permanent magnets is that the field strength cannot be controlled. Yet, the mechanical implementation as used here allows for an easy and reproducible replacement of the magnets with weaker or stronger magnets. This may need to be considered for species that have a much different mass over magnetic moment ratio than the species used here, or if a smaller transverse velocity spread is required.

The solenoids consist of copper capillary of 1.5 mm outer diameter and 0.6 mm inner diameter, with four and a half windings around a bore of 3 mm in diameter. A winding process was developed that guarantees that all 100 solenoids have the same dimensions, with little to no deviation. The capillary is wound on a rod with evenly spaced grooves that ensure equal spacing between the windings. The straight entrance and exit sections of the solenoid capillary act as feed-through in the aluminium flange, and define the correct positioning of the solenoid with respect to the beam axis. Sufficient accuracy is obtained with a precise mold, in which a rod is fixed through the solenoid at the correct distance from the aluminium flange. A polyvinyl chloride (PVC) bus is used to isolate the capillary from the aluminium flange. Torr-seal is used to obtain a seal suitable for high-vacuum. This resin is heated and applied from both sides and stirred until no air bubbles remain. The accuracy of the solenoid positions and rotations relative to the aluminium mounts is tested by installing them in a decelerator module, and inserting a 2.8-mm diameter PVC rod through the centers of the 20 solenoids. When extra friction is detected during this test, the offending solenoid is identified and replaced.  When the rod is transmitted smoothly, the solenoids are accepted.

The hollow capillary allows for a flow of cooling liquid that is thus in direct contact with the solenoids inside the vacuum chamber. The cooling liquid is pumped from a single source (ThermoFisher system I heat exchanger), and a series of manifolds is used to distribute the flow equally, with silicon tubes running to and from the solenoids. These tubes are connected directly to the solenoid feedthroughs, using brass rings that provide a tight seal. Initially automobile coolant (PROFCLEAN, all-season) was used, but it proved unsuitable for the narrow solenoid capillary. The anti-freeze made it too viscous and it occasionally caused obstructions. We switched to racing coolant (BARDAHL ref. 13113), which forgoes the anti-freeze agent but still has anti-corrosive properties.

For optimal cooling, a rapid flow is needed, which depends on the pressure difference between the liquid at the entrance and exit of the solenoids. The maximum pressure difference before the silicon tubes would detach from the solenoids was found to be 1.5 bar. In the experiments, a pressure difference of 0.8 bar is used, which has proven sufficient to run the solenoids at repetition rates of 10 Hz.

In the early testing stages, when applying current pulses to the solenoids, they would audibly flex due to the self-inductive force on the capillary. In some cases, after a few months of operation, the capillary was found to tear open and spill coolant. To prevent breaks, solenoids were coated with Torr-seal to combat the flexing motion. After being applied to the solenoid, the resin is heated to reduce viscosity and release any trapped air. The solenoid is rotated until the resin hardens, to prevent the resin from accumulating on one side. A Teflon rod is inserted into the center of the solenoid when applying the resin, such that the bore of the solenoid remains fully unobstructed. Since Torr-seal does not adhere to Teflon, the rod is easily removed after the resin has set. No further ruptures of solenoids were observed, for almost non-stop operation during a full year.

When the five modules were fully assembled with solenoids and hexapoles, a final check was performed to confirm the precise alignment between the modules. While very small deviations between the modules are expected to be compensated by the hexapoles, larger deviations could significantly reduce the number of transmitted particles. However, with a theodolite we observed a straight sight-line through the entire decelerator, which is a good indication that the modules are well aligned.

\subsection{Electronics}\label{sec:elec}

Each solenoid is connected to an individual PCB that is powered by a separate 24-volt power supply. These low voltages are sufficient because of the low resistance in the solenoids ($\approx$ 1 m$\Omega$), and allow for the use of relatively cheap FET-based electronic components compared to the high-voltage insulated gate bipolar transistor (IGBT) switches that are used in Zeeman decelerators of other design. The PCBs can provide two successive current pulses of up to 4.5 kA per deceleration sequence. With these currents, magnetic fields above 2 T are generated in the solenoid center. The currents are pulsed for 10-100 $\mu$s, depending on the mean velocity of the molecular packet. The duration of these pulses, in combination with the maximum current and the experimental repetition rate, defines the amount of heat dissipated in the solenoid and in the PCB. In order to maintain repetition rates of at least 10 Hz, the PCB is liquid cooled using the same manifold that provides the cooling liquid to the solenoids. A 40-mm fan is mounted on each PCB for additional cooling.

\begin{figure}[!htb]
\centering
\resizebox{1.0\linewidth}{!}
{\includegraphics{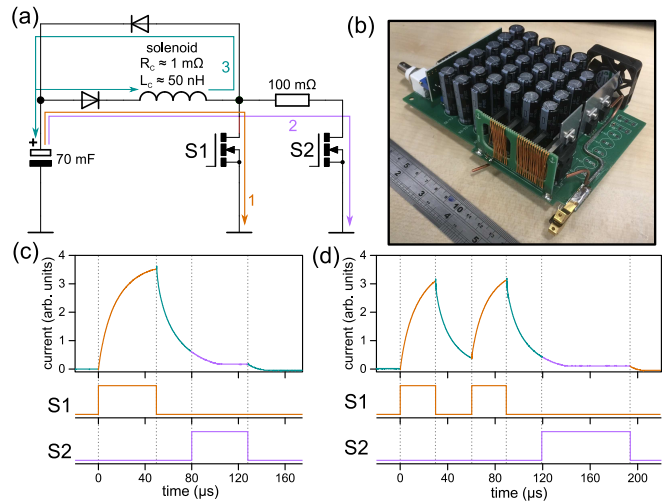}}
\caption{Schematic representation of the driving electronics and the current pulse through a solenoid. (\emph{a}) A schematic drawing of the electronic circuit enabling the solenoid current. The capacitor of 70 mF in the bottom-left represents a parallel array of 32 smaller capacitors of 2.2 mF. The PCB in the photograph of panel (\emph{b}) features these capacitors as the main component. Current profile of a single (\emph{c}) or double (\emph{d}) pulse through the solenoid derived from the voltage induced in a smaller pickup solenoid. The trigger pulses for gates S1 and S2 to generate these pulses are shown underneath the current profiles. The color in the profiles correspond to the currently active electronic pathway in panel (\emph{a}). The colored pathways correspond to: 1. maximum current through solenoid, 2. reduced current through solenoid, and 3. dissipation of any leftover current in the solenoid.}
\label{fig:electro}
\end{figure}

Fig. \ref{fig:electro}(a) shows a simplified schematic of the electronic circuit of the PCB, including the solenoid. The schematic circuit also shows the three possible pathways in which the current can flow. The first pathway is active when gate S1 is closed, and is used to drive the capacitor charge through the solenoid, providing up to 4.5 kA of current for a short time. The second pathway is engaged when S2 is closed, which adds a 100 m$\Omega$ resistance in series with the solenoid. With this resistor in series, the current through the solenoid is severely reduced, down to about 150 A. This small current is sufficient to provide a bias magnetic field inside the solenoids that prevent Majorana transitions in the molecular packet \cite{Cremers:PRA95:043415}. This pathway is used when the particles pass through the solenoid after switch S1 has opened. The third pathway is used to eliminate flyback when gates S1 and S2 are open, by sending any excess charge back to the capacitors.

Depending on the mode of operation, switch S1 is opened in a single (deceleration and acceleration mode) or double pulse (hybrid mode). Fig. \ref{fig:electro}(c) and (d) show the current profiles through the solenoid when it receives a single and double pulse, respectively. These profiles were obtained from the induced voltage over a miniature solenoid that was placed inside the center of a decelerator solenoid. This provided us with the shape of the current pulses in our solenoids. As can be seen in the measured profiles, there is a significant rise and fall time in the current profiles. This is due to the self-inductance in the solenoid (50 nH) and surrounding electronics. In previous versions, the rise time was reduced by using a fourth pathway that increased the potential difference through the solenoid to 60 V. However, the reduction in rise time was too small to justify the increased stress on the switching transistors. The rise and fall profiles are necessarily taken into account when calculating the switching times of the solenoids for a deceleration sequence. The second pathway is automatically engaged 30 $\mu$s after the single or double S1 pulse is terminated, and remains active during 50 - 80 $\mu$s. The 30-$\mu$s delay allows the current to decay more quickly through the second pathway and prevents excessive heat to be generated in the resistor. The 50 - 80 $\mu$s duration was chosen as sufficient time for all molecules in a packet to benefit from the quantization field.

One of the challenges in designing the PCB is the relatively high current that flows through the components, which causes a large amount of heat dissipation. Therefore, where possible the tracks between electrical components are kept at low resistance by using as much of the PCB surface as possible. Additionally, cooling liquid is flowed through copper capillary that is directly soldered onto the PCB area that experiences the most heat dissipation. The connection between PCB and solenoid is necessarily kept as short as physically possible, by attaching it directly onto the solenoid legs with a set of brass clamps. These brass clamps and the copper cooling tubes on the PCB are seen in the bottom-right of Fig. \ref{fig:electro}(b). The length of the solenoid leg between the clamp and the beginning of the solenoid is only about 23 mm, yet it constitutes about 40\% of the total current carrying wire. For the 100 m$\Omega$ resistance in the second current pathway a conventional resistor proved unsuitable, as its thermal conductivity was not sufficient to disperse the heat pulse generated in a single switching sequence. Instead a bifilar coil of copper wire is used as a resistor. The wire is wrapped around a rectangular grid that maximizes the surface area. This component is custom made, and can be seen next to the brass clamps in Fig \ref{fig:electro}(b).

The switching timings for the solenoids are calculated using a molecular trajectory program, by simulating a synchronous particle. The list of timings is sent to a small computer (Raspberry Pi), which passes them onto a timing PCB for each solenoid. Each individual timing is addressed with a unique number that is recognized by a dual in-line package (DIP)-switch system on the PCBs. Once loaded, these timing PCBs are synchronized to a single 20 MHz clock, and start counting when the system receives an external trigger. At the loaded pulse timings, they send a trigger to the PCB connected to the solenoid, which then generates a pulse for the appropriate duration.

\subsection{Transverse focusing}\label{sec:trans}
\begin{figure}[!tb]
\centering
\resizebox{1.0\linewidth}{!}
{\includegraphics{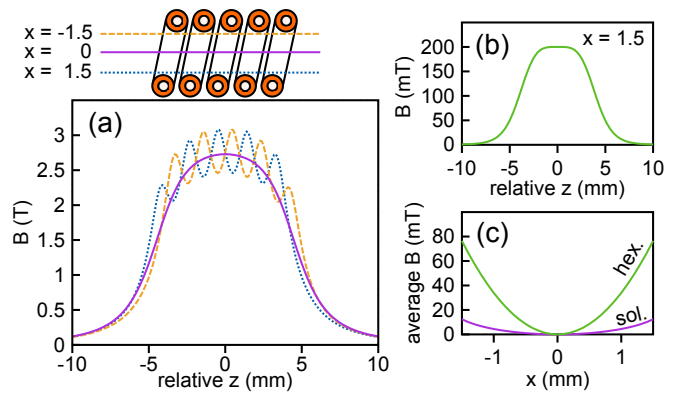}}
\caption{(\emph{a}) The magnetic field strength, $B$, inside one of our solenoids running 4.5 kA of current, as a function of the longitudinal position, $z$, relative to the solenoid center. A cross section of a solenoid is shown above the graph, with the lines along which the magnetic fields were calculated. (\emph{b}) The magnetic field strength of a hexapole used in our experiments, at a radial offset of 1.5 mm, as a function of the relative longitudinal position. (\emph{c}) The average magnetic field contribution of the solenoids (purple) and hexapoles (green) along the transverse coordinate, $x$. The average was taken over two periods of the active decelerator (see equation \ref{eq:avgper} and the accompanying text).}
\label{fig:Fieldcompare}
\end{figure}

The magnetic field inside a solenoid running 4.5 kA of current was calculated by applying the Biot-Savart law to a solenoid geometry that matched our solenoid design. The result is shown in Fig. \ref{fig:Fieldcompare}(a). The fields are calculated along lines parallel to the beamline at three different transverse positions. The magnetic field distributions close to the inside of the solenoid are shown to fluctuate based on the proximity to the nearest capillary. Consequently, the entrance and exit connections of a solenoid cause an asymmetry in the induced magnetic fringe fields near the ends of a solenoid. Adjacent solenoids are therefore rotated such that all entrance connections are positioned on alternating sides of the beam axis, to average out the asymmetry. For LFS particles, the average transverse gradient of the solenoids provides a focusing force directly inside the solenoids, but results in a small defocusing force in the fringe fields.

The magnetic field inside a hexapole was calculated using Radia 4.29 \cite{RADIA_4.29}, and is shown in Fig. \ref{fig:Fieldcompare}(b), for a transverse offset of 1.5 mm. On-axis the hexapole magnetic field is by definition zero, resulting in a magnetic field gradient that is purely focusing for all LFS particles. From a quick comparison of the transverse gradients of the solenoid and hexapole fields, it would seem that the solenoids provide the stronger transverse focusing. However, due to the rising and falling of the currents inside the solenoids, this is not the case. We show this with a particle simulation that uses the routines from a previous work \cite{Cremers:PRA95:043415}, but with updated current profiles and field distributions. We simulate the synchronous particle to obtain the time-dependent magnetic fields, and calculate the average contribution of the solenoids and hexapoles to the field as a function of the transverse coordinate, $x$. The field was averaged over two time periods of the decelerator to average out the effect of the solenoid fringe fields:
\begin{equation}
B_{avg}(x) = \frac{1}{2T}\int_{0}^{2T} B(z(t),x, t) dt
\label{eq:avgper}
\end{equation}
where one period, $T$, is the time in which the synchronous particle passes one solenoid and one hexapole, and $z$ is the longitudinal position of the synchronous particle. $B$ is the magnetic field strength inside the decelerator, which depends on the position and the time in the decelerator. The vertical offset, $y$, is set to zero. We chose a decelerator sequence that is representative for the experiments described in section \ref{sec:res}, specifically a guiding sequence at 550~m/s. The result of this simulation can be seen in Fig\ref{fig:Fieldcompare}(c), for the contributions of the solenoids and hexapoles separately. In the solenoid contribution, the minimum magnetic field was set to zero, to show only the transverse gradient. The permanent magnetic hexapoles contribute significantly more to the transverse focusing of the particle beam, about five times more compared to the solenoids. By contrast, the solenoids affect the longitudinal velocity components, much more than the hexapoles. These two elements together allow for nearly independent control over the longitudinal and transverse properties of decelerated particles. The transverse and longitudinal focusing properties are also independent of the length of the decelerator, due to the phase-stable nature of the decelerator \cite{Cremers:PRA98:033406}.

\subsection{Experimental setup}\label{sec:setup}

\begin{figure}[!htb]
\centering
\resizebox{1.0\linewidth}{!}
{\includegraphics{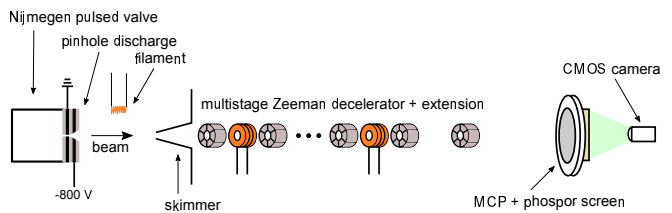}}
\caption{Schematic diagram of the experimental setup. The extension consists of eight hexapoles and is positioned between the exit of the decelerator and the detection region. The distance from the valve to the skimmer is approximately 100 mm, and the distance from the final hexapole to the MCP detector is 250 mm.}
\label{fig:setup}
\end{figure}

In order to characterize the Zeeman decelerator, and experimentally validate the suitability of the solenoid-hexapole concept for Zeeman deceleration, beams of atoms or molecules are passed through the Zeeman decelerator and their arrival in the detection area is monitored. Control over the longitudinal velocity of oxygen atoms and molecules was demonstrated in a previous paper \cite{Cremers:PRA98:033406}, where we presented time-of-flight (TOF) profiles of the atoms and molecules exiting the decelerator. Excellent agreement between the observed and simulated time-of-flight profiles was found. In this paper, we primarily focus on the transverse properties of the decelerator. For this, we generate beams of metastable rare gas atoms (He, Ne, Ar and Kr), and guide them through the decelerator in hybrid mode. We measure the radial distribution of the atoms upon exiting the Zeeman decelerator, which probes the transverse velocity spread of the Zeeman decelerated packet. A schematic representation of the experimental setup can be seen in Fig. \ref{fig:setup}.

Rare gas atoms are brought into a metastable state via an electric discharge, by promoting one electron from the highest occupied atomic orbital to the lowest unoccupied orbital. Rare gas atoms in these metastable states have enough internal energy to release an Auger electron when they impinge on a microchannel plate (MCP) detector \cite{Dunning:EMPS29B}. The subsequent electron avalanche can be measured directly. In addition, a phosphor screen is attached to the MCP that generates a flash of light at the impact position. A complementary metal-oxide semiconductor (CMOS) camera is used to record the light from the phosphor screen to retrieve a spatial map of the metastable atoms exiting the decelerator.

A Nijmegen pulsed valve is used to create supersonic beams of rare gas atoms \cite{Yan:RSI84:023102}. The gases are expanded at room temperature with a repetition rate of 10 Hz. The atoms are brought into the metastable states by a discharge in a pinhole nozzle, similar to the one used by Ploenes et al. \cite{Ploenes:RSI87:053305}. The front plate of the discharge nozzle is set to -800 V for 30-50 microsecond pulses, depending on the gas. A hot filament, consisting of a 0.3-mm diameter tungsten wire in a tight spiral through which 3 A of current is passed, is mounted near the valve orifice and used to stabilize the discharge process. The metastable atoms pass through a skimmer and into the multistage Zeeman decelerator. Between the exit of the decelerator and the detection region, an extension consisting of eight hexapoles is placed. The MCP with phosphor screen is placed 250 mm from the final hexapole. A voltage difference of 1600 V is applied over two MCP plates, and the phosphor screen is set to 4500 V. The MCP voltage is pulsed using a Behlke high-voltage switch, limiting the measurement time to only 20 microseconds, which is sufficient to distinguish the Zeeman decelerated atoms from the remainder of the gas pulse.

\begin{table*}[!htb]
\centering
\def\arraystretch{1.3}
\begin{tabular}{|c|c|c|c|c|c|}
\hline
atom & state & m/$\mu$ (amu/$\mu_B$) &  exp. $\bar{v}_z$ (m/s) & selected $\bar{v}_z$ (m/s) & max. $v_x$ (m/s) for 90\% of packet \\
\hline
He & 1s2s $^3$S$_1$ & 2.00  & 1950 & 1750 & 12.2 \\
Ne & 2p$^5$3s $^3$P$_0$ & 6.73  & 1210 & 1100 & 8.5 \\
Ar & 3p$^5$4s $^3$P$_0$ & 13.32 & 790 & 790 & 6.0 \\
Kr & 4p$^5$5s $^3$P$_0$ & 27.93 & 550 & 550 & 3.5 \\
\hline
\end{tabular}
\caption{Various properties of metastable rare gas atom beams that are relevant for Zeeman deceleration. $v_z$ denotes the longitudinal velocity, and $v_x$ the transverse velocity.}
\label{tab:one}
\end{table*}

\section{Results}\label{sec:res}

\begin{figure}[!b]
\centering
\resizebox{1.0\linewidth}{!}
{\includegraphics{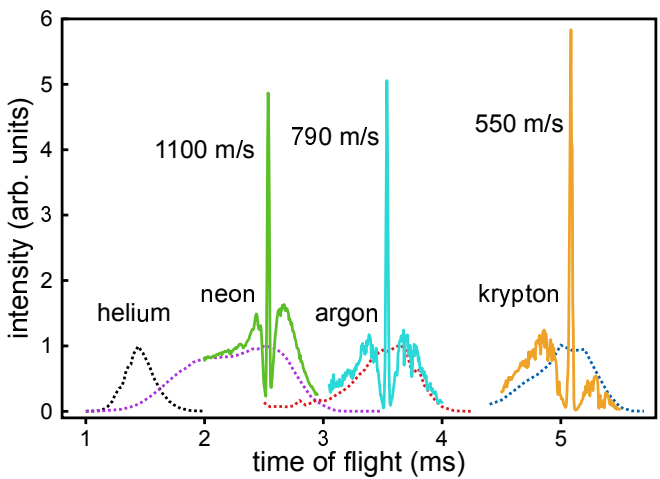}}
\caption{Experimental TOF profiles of metastable rare gas atoms, after passing through the multistage Zeeman decelerator. Profiles with dashed lines correspond to beams passed through the decelerator when the solenoids were not activated. For neon, argon and krypton, the decelerator is used in hybrid mode to guide the atoms through the decelerator at constant speed as indicated, resulting in the profiles with solid lines. The relative intensities for profiles corresponding to different rare gas beams are normalized with respect to the dashed curves.}
\label{fig:rg_guide}
\end{figure}

Time-of-flight profiles, in combination with radial distributions were measured for metastable He, Ne, Ar and Kr atoms that exit the Zeeman decelerator. The electronic states together with the ratios of mass to magnetic dipole moment are listed in table \ref{tab:one}. To determine the mean velocity of the metastable atom beam, TOF profiles were first measured for all species with the solenoids off, i.e., the atoms are then only transversally focused by the hexapole array. The resulting profiles are shown in Fig. \ref{fig:rg_guide} as the dashed traces, which are all normalized with respect to each other. From these profiles, and the known discharge ignition time and travel path from the source to the detector, the mean beam speeds as given in table \ref{tab:one} were deduced. These speeds are significantly higher than what may be expected for a room temperature expansion, which is attributed to heating effects due to the discharge process.

\begin{figure}[!htb]
\centering
\resizebox{1.0\linewidth}{!}
{\includegraphics{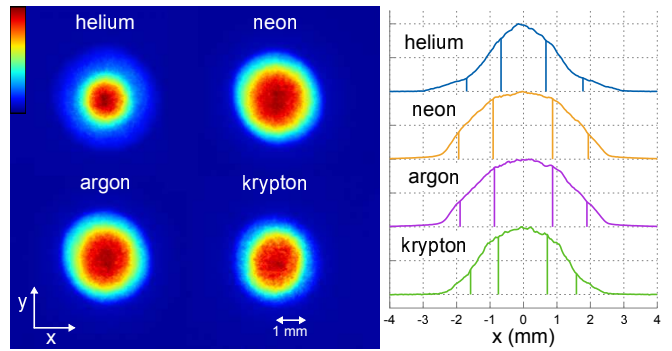}}
\caption{Experimental spatial distributions in the transverse plane of packets of metastable He, Ne, Ar and Kr atoms exiting the multistage Zeeman decelerator. The false-color scale depicts the intensity of the recorded signal. The graphs on the right depict the radial cuts, with different vertical offset and normalized intensity. The inner and outer vertical lines in the graphs encapsulate 50\% and 90\% of the total number of atoms in the packet, respectively. (color online)}
\label{fig:raddis}
\end{figure}

\begin{figure*}[!htb]
\centering
\resizebox{1.0\linewidth}{!}
{\includegraphics{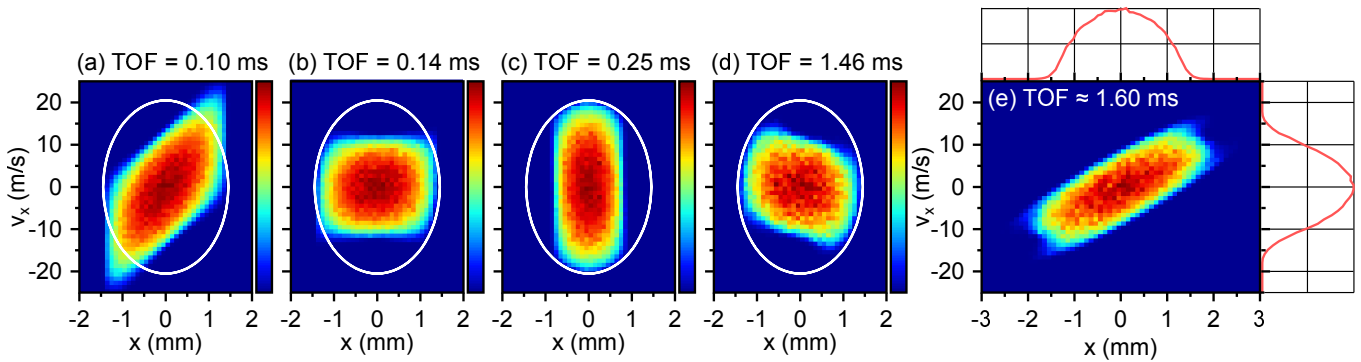}}
\caption{Simulated transverse phase-space diagrams of metastable helium at different times-of-flight in our multistage Zeeman decelerator with the solenoids off. The atom density is displayed with a false-color scale. The diagrams (a)-(d) include the separatrix of stable transverse phase space as a white ellipse. The diagrams were generated at: (a) the entrance of the decelerator (b) the first 'velocity node' (c) the first 'spatial node' (d) the end of the hexapole extension (e) the position of the MCP detector in the experiment. For this diagram the spatial and velocity distributions are also shown separately. (color online)}
\label{fig:tphase}
\end{figure*}

The Zeeman decelerator is subsequently used in hybrid mode and programmed to select a velocity that is close to the mean velocity of the beam. This was possible for all metastable atoms, except for the helium beam which traveled too fast to be significantly affected by the pulsed magnetic fields. The measured TOF profiles after guiding for Ne, Ar and Kr are shown in Fig. \ref{fig:rg_guide}. The central narrow peak in the TOF profiles contains the packet of guided atoms, and is well separated from the remainder of the atom beam even for beams of metastable Ne atoms that travel with a mean speed exceeding 1000 m/s. The guided packets of Ne, Ar and Kr had a similar longitudinal velocity spread, approximately 20 m/s, and spatial spreads of around 10 mm, 13 mm and 16 mm, respectively. The ability of the decelerator to manipulate the longitudinal motion of paramagnetic particles has been explored in a previous work \cite{Cremers:PRA98:033406}. Here, we focus on the transverse distributions of atoms at the end of the decelerator.

The radial distributions were measured for the selected packets of Ne, Ar and Kr, by gating the MCP such that only the guided packets are recorded. For helium an arrival time was chosen that corresponded with a forward velocity of about 1750 m/s. The images that result from measurements averaged over 50 shots of the experiment are shown in Fig. \ref{fig:raddis}. A false-color scale is used to display the intensity of the signal, of which the scale is seen in the top-left. Normalized cross sections of the radial distributions are displayed in the graph on the right. Vertical lines indicate the part of the packet that contains 90\% (outer vertical lines) and 50\% (inner vertical lines) of the atoms within the packet.

The spatial distribution observed in the images result from spatial spread of the packet at the exit of the hexapole extension, in combination with the transverse spread that is built up during the free flight towards the detector. The latter is governed by the inherent transverse velocity distribution in the packet, which in turn is governed by the transverse focusing forces of the hexapole array in the Zeeman decelerator. Since the hexapole magnets induce a magnetic field that is quadratic in the radial position, the atoms move as a harmonic oscillator in the transverse directions. In this motion, the maximum transverse velocity in the decelerator is found on the beam axis. By calculating the free-flight time of the atoms to the detector, and relating this to the maximum deviation a particle reaches when it originates from the beam axis, an upper limit to the radial velocity can be deduced from the images. Using this method, the maximum transverse velocities are calculated for the part of the packet containing 90\% of the guided atoms. The resulting values are given in table \ref{tab:one}, and are seen to approximately scale with $\sqrt{m/\mu}$ as expected for a harmonic oscillator.

The largest deviation from this expected result is found in the helium velocity, which should be around 15.6 m/s. This deviation stems from the measured radial distribution, which is more narrow than a calculation of the magnetic forces would suggest. More numerical trajectory simulations were performed to look directly at the transverse phase-space distributions of the metastable helium beam inside the decelerator. The results are shown in Fig. \ref{fig:tphase}. In the simulation we generated an initial helium distribution that is an approximation of the beam distribution in the experiment. While it is not an exact match, it can be used to give an understanding of the phase-space evolution in the decelerator. Due to the aperture of the skimmer, the helium distribution is truncated when flown into the first hexapole, as seen in the Fig. \ref{fig:tphase}(a). With the average magnetic field gradients calculated in section \ref{sec:trans}, the transverse separatrix for helium atoms in our decelerator was calculated, and displayed over the phase spaces. Because the separatrix is not uniformly filled, the transverse phase-space distributions show significant changes in time. The particles make around five harmonic oscillations before they reach the final hexapole. The transverse phase-space distribution when the helium exits the final hexapole is shown in Fig. \ref{fig:tphase}(d). Because the particles are close to a 'velocity node' at this time, the velocity distribution is more narrow than would be expected considering the separatrix. This is reflected in the spatial distribution found at the MCP detector position, seen in Fig. \ref{fig:tphase}(e). By changing the length of the decelerator, the separatrix would remain unchanged, but the helium particles would likely be in a different part of their transverse oscillation and therefore show a greater spread when detected.

For Ne, Ar and Kr, the radial distributions were also measured when the solenoids in the Zeeman decelerator were not used and switched off (data not shown). These distributions showed no significant differences from the images recorded when the Zeeman decelerator was used in hybrid mode to guide the beams at constant speed. This substantiates the design principle of the decelerator, which was chosen to have transverse focusing independent of the longitudinal motion. We therefore conclude that the hexapoles focus the beam an equal amount regardless of the solenoid magnetic fields, i.e., the transverse properties of the Zeeman decelerated packets are governed by the hexapole array.

\section{Summary and Outlook}\label{sec:sum}
In this paper, a detailed account is given of the implementation of a multistage Zeeman decelerator that is specifically designed for crossed beams scattering experiments. This Zeeman decelerator is conceptually different from other multistage Zeeman decelerators. It consists of an alternating array of solenoids and magnetic hexapoles that decouple the longitudinal and transverse properties of the decelerator. The solenoid design allows for easy in-vacuum positioning while facilitating efficient internal liquid cooling. In addition, the low resistance and self induction of the solenoid allows for the use of cost-effective low-voltage and FET-based electronic components.

The decelerator features a modular design, that allows for the easy addition or removal of individual modules. The modules are mechanically coupled and aligned with respect to each other with high precision, and without breaking the solenoid-hexapole array. Excellent vacuum properties are maintained by sufficient pumping capacity delivered by turbo pumps that are directly connected to the modules, and the solenoids and printed-circuit boards are internally cooled. In principle, this allows for operation of the Zeeman decelerator at repetition rates exceeding 10 Hz; the upper limit to the repetition rate depends on the duration of the current pulses in the sequence.

The transverse focusing properties of the hexapoles are investigated by recording the radial distributions of different metastable rare-gas beams that are passed through the decelerator. When the hexapoles are used, the radial distributions are found to be unaffected by the magnetic fields of the solenoids, confirming that the hexapoles govern the transverse motion of particles inside the decelerator. The observed spatial distributions probe the transverse velocity distributions of the Zeeman decelerated beams, which were found to be in good agreement with the predictions from numerical calculations of the deceleration process.

The decelerator presented here is ideally suited for crossed beams scattering experiments. It can easily be extended with a module that interfaces the decelerator with an interaction region where the Zeeman-decelerated beam is intercepted by a secondary beam. The design also allows for the addition of custom-made modules that, for instance, provide additional optical access near the exit of the decelerator. One may also envision the addition of a curved module in which adjacent solenoids and hexapoles are positioned under a small angle. This would allow for efficient removal of carrier gas atoms from the beam which is important for collision experiments involving traps or surfaces. A curved extension may also allow for the implementation of merged beam geometries to reach (ultra)low collision energies well below 1 Kelvin \cite{Henson:Science338:234}.

\section{Acknowledgments}
The research leading to these results has received funding from the European Research Council under the European Union's Seventh Framework Programme (FP7/2007-2013/ERC grant agreement nr. 335646 MOLBIL). This work is part of the research program of the Netherlands Organization for Scientific Research (NWO). We thank Gerben Wulterkens for the mechanical design of the multistage Zeeman decelerator.



\begin{thebibliography}{29}
\expandafter\ifx\csname natexlab\endcsname\relax\def\natexlab#1{#1}\fi
\expandafter\ifx\csname bibnamefont\endcsname\relax
  \def\bibnamefont#1{#1}\fi
\expandafter\ifx\csname bibfnamefont\endcsname\relax
  \def\bibfnamefont#1{#1}\fi
\expandafter\ifx\csname citenamefont\endcsname\relax
  \def\citenamefont#1{#1}\fi
\expandafter\ifx\csname url\endcsname\relax
  \def\url#1{\texttt{#1}}\fi
\expandafter\ifx\csname urlprefix\endcsname\relax\def\urlprefix{URL }\fi
\providecommand{\bibinfo}[2]{#2}
\providecommand{\eprint}[2][]{\url{#2}}

\bibitem[{\citenamefont{Lee}(1987)}]{Lee:Science236:793}
\bibinfo{author}{\bibfnamefont{Y.~T.} \bibnamefont{Lee}},
  \bibinfo{journal}{Science} \textbf{\bibinfo{volume}{236}},
  \bibinfo{pages}{793} (\bibinfo{year}{1987}), ISSN \bibinfo{issn}{0036-8075}.

\bibitem[{\citenamefont{DePaul et~al.}(1993)\citenamefont{DePaul, Pullman, and
  Friedrich}}]{Depaul:JPC97:2167}
\bibinfo{author}{\bibfnamefont{S.}~\bibnamefont{DePaul}},
  \bibinfo{author}{\bibfnamefont{D.}~\bibnamefont{Pullman}}, \bibnamefont{and}
  \bibinfo{author}{\bibfnamefont{B.}~\bibnamefont{Friedrich}},
  \bibinfo{journal}{The Journal of Physical Chemistry}
  \textbf{\bibinfo{volume}{97}}, \bibinfo{pages}{2167} (\bibinfo{year}{1993}).

\bibitem[{\citenamefont{Chefdeville et~al.}(2013)\citenamefont{Chefdeville,
  Kalugina, van~de Meerakker, Naulin, Lique, and
  Costes}}]{Chefdeville:Science341:06092013}
\bibinfo{author}{\bibfnamefont{S.}~\bibnamefont{Chefdeville}},
  \bibinfo{author}{\bibfnamefont{Y.}~\bibnamefont{Kalugina}},
  \bibinfo{author}{\bibfnamefont{S.~Y.~T.} \bibnamefont{van~de Meerakker}},
  \bibinfo{author}{\bibfnamefont{C.}~\bibnamefont{Naulin}},
  \bibinfo{author}{\bibfnamefont{F.}~\bibnamefont{Lique}}, \bibnamefont{and}
  \bibinfo{author}{\bibfnamefont{M.}~\bibnamefont{Costes}},
  \bibinfo{journal}{Science} \textbf{\bibinfo{volume}{341}},
  \bibinfo{pages}{1094} (\bibinfo{year}{2013}).

\bibitem[{\citenamefont{Henson et~al.}(2012)\citenamefont{Henson, Gersten,
  Shagam, Narevicius, and Narevicius}}]{Henson:Science338:234}
\bibinfo{author}{\bibfnamefont{A.~B.} \bibnamefont{Henson}},
  \bibinfo{author}{\bibfnamefont{S.}~\bibnamefont{Gersten}},
  \bibinfo{author}{\bibfnamefont{Y.}~\bibnamefont{Shagam}},
  \bibinfo{author}{\bibfnamefont{J.}~\bibnamefont{Narevicius}},
  \bibnamefont{and}
  \bibinfo{author}{\bibfnamefont{E.}~\bibnamefont{Narevicius}},
  \bibinfo{journal}{Science} \textbf{\bibinfo{volume}{338}},
  \bibinfo{pages}{234} (\bibinfo{year}{2012}).

\bibitem[{\citenamefont{Osterwalder}(2015)}]{Osterwalder:EPJ-TI2:10}
\bibinfo{author}{\bibfnamefont{A.}~\bibnamefont{Osterwalder}},
  \bibinfo{journal}{EPJ Techniques and Instrumentation}
  \textbf{\bibinfo{volume}{2}}, \bibinfo{pages}{10} (\bibinfo{year}{2015}).

\bibitem[{\citenamefont{Stuhl et~al.}(2014)\citenamefont{Stuhl, Hummon, and
  Ye}}]{Stuhl:ARPC65:501}
\bibinfo{author}{\bibfnamefont{B.~K.} \bibnamefont{Stuhl}},
  \bibinfo{author}{\bibfnamefont{M.~T.} \bibnamefont{Hummon}},
  \bibnamefont{and} \bibinfo{author}{\bibfnamefont{J.}~\bibnamefont{Ye}},
  \bibinfo{journal}{Ann. Rev. Phys. Chem.} \textbf{\bibinfo{volume}{65}},
  \bibinfo{pages}{501} (\bibinfo{year}{2014}).

\bibitem[{\citenamefont{Brouard et~al.}(2014)\citenamefont{Brouard, Parker, and
  van~de Meerakker}}]{Brouard:CSR43:7279}
\bibinfo{author}{\bibfnamefont{M.}~\bibnamefont{Brouard}},
  \bibinfo{author}{\bibfnamefont{D.~H.} \bibnamefont{Parker}},
  \bibnamefont{and} \bibinfo{author}{\bibfnamefont{S.~Y.~T.}
  \bibnamefont{van~de Meerakker}}, \bibinfo{journal}{Chem. Soc. Rev.}
  \textbf{\bibinfo{volume}{43}}, \bibinfo{pages}{7279} (\bibinfo{year}{2014}).

\bibitem[{\citenamefont{van~de Meerakker et~al.}(2012)\citenamefont{van~de
  Meerakker, Bethlem, Vanhaecke, and Meijer}}]{Meerakker:CR112:4828}
\bibinfo{author}{\bibfnamefont{S.~Y.~T.} \bibnamefont{van~de Meerakker}},
  \bibinfo{author}{\bibfnamefont{H.~L.} \bibnamefont{Bethlem}},
  \bibinfo{author}{\bibfnamefont{N.}~\bibnamefont{Vanhaecke}},
  \bibnamefont{and} \bibinfo{author}{\bibfnamefont{G.}~\bibnamefont{Meijer}},
  \bibinfo{journal}{Chemical Reviews} \textbf{\bibinfo{volume}{112}},
  \bibinfo{pages}{4828} (\bibinfo{year}{2012}).

\bibitem[{\citenamefont{Bethlem et~al.}(1999)\citenamefont{Bethlem, Berden, and
  Meijer}}]{Bethlem:PRL83:1558}
\bibinfo{author}{\bibfnamefont{H.~L.} \bibnamefont{Bethlem}},
  \bibinfo{author}{\bibfnamefont{G.}~\bibnamefont{Berden}}, \bibnamefont{and}
  \bibinfo{author}{\bibfnamefont{G.}~\bibnamefont{Meijer}},
  \bibinfo{journal}{Phys. Rev. Lett.} \textbf{\bibinfo{volume}{83}},
  \bibinfo{pages}{1558} (\bibinfo{year}{1999}).

\bibitem[{\citenamefont{Vanhaecke et~al.}(2007)\citenamefont{Vanhaecke, Meier,
  Andrist, Meier, and Merkt}}]{Vanhaecke:PRA75:031402}
\bibinfo{author}{\bibfnamefont{N.}~\bibnamefont{Vanhaecke}},
  \bibinfo{author}{\bibfnamefont{U.}~\bibnamefont{Meier}},
  \bibinfo{author}{\bibfnamefont{M.}~\bibnamefont{Andrist}},
  \bibinfo{author}{\bibfnamefont{B.~H.} \bibnamefont{Meier}}, \bibnamefont{and}
  \bibinfo{author}{\bibfnamefont{F.}~\bibnamefont{Merkt}},
  \bibinfo{journal}{Phys. Rev. A} \textbf{\bibinfo{volume}{75}},
  \bibinfo{pages}{031402(R)} (\bibinfo{year}{2007}).

\bibitem[{\citenamefont{Narevicius et~al.}(2007)\citenamefont{Narevicius,
  Parthey, Libson, Narevicius, Chavez, Even, and Raizen}}]{Narevicius:NJP9:358}
\bibinfo{author}{\bibfnamefont{E.}~\bibnamefont{Narevicius}},
  \bibinfo{author}{\bibfnamefont{C.~G.} \bibnamefont{Parthey}},
  \bibinfo{author}{\bibfnamefont{A.}~\bibnamefont{Libson}},
  \bibinfo{author}{\bibfnamefont{J.}~\bibnamefont{Narevicius}},
  \bibinfo{author}{\bibfnamefont{I.}~\bibnamefont{Chavez}},
  \bibinfo{author}{\bibfnamefont{U.}~\bibnamefont{Even}}, \bibnamefont{and}
  \bibinfo{author}{\bibfnamefont{M.~G.} \bibnamefont{Raizen}},
  \bibinfo{journal}{New J. of Physics} \textbf{\bibinfo{volume}{9}},
  \bibinfo{pages}{358} (\bibinfo{year}{2007}).

\bibitem[{\citenamefont{Gilijamse et~al.}(2006)\citenamefont{Gilijamse,
  Hoekstra, van~de Meerakker, Groenenboom, and
  Meijer}}]{Gilijamse:Science313:1617}
\bibinfo{author}{\bibfnamefont{J.~J.} \bibnamefont{Gilijamse}},
  \bibinfo{author}{\bibfnamefont{S.}~\bibnamefont{Hoekstra}},
  \bibinfo{author}{\bibfnamefont{S.~Y.~T.} \bibnamefont{van~de Meerakker}},
  \bibinfo{author}{\bibfnamefont{G.~C.} \bibnamefont{Groenenboom}},
  \bibnamefont{and} \bibinfo{author}{\bibfnamefont{G.}~\bibnamefont{Meijer}},
  \bibinfo{journal}{Science} \textbf{\bibinfo{volume}{313}},
  \bibinfo{pages}{1617} (\bibinfo{year}{2006}).

\bibitem[{\citenamefont{von Zastrow et~al.}(2014)\citenamefont{von Zastrow,
  Onvlee, Vogels, Groenenboom, van~der Avoird, and van~de
  Meerakker}}]{Zastrow:NatChem6:216}
\bibinfo{author}{\bibfnamefont{A.}~\bibnamefont{von Zastrow}},
  \bibinfo{author}{\bibfnamefont{J.}~\bibnamefont{Onvlee}},
  \bibinfo{author}{\bibfnamefont{S.~N.} \bibnamefont{Vogels}},
  \bibinfo{author}{\bibfnamefont{G.~C.} \bibnamefont{Groenenboom}},
  \bibinfo{author}{\bibfnamefont{A.}~\bibnamefont{van~der Avoird}},
  \bibnamefont{and} \bibinfo{author}{\bibfnamefont{S.~Y.~T.}
  \bibnamefont{van~de Meerakker}}, \bibinfo{journal}{Nat. Chem.}
  \textbf{\bibinfo{volume}{6}}, \bibinfo{pages}{216} (\bibinfo{year}{2014}).

\bibitem[{\citenamefont{Onvlee et~al.}(2016)\citenamefont{Onvlee, Vogels, and
  van~de Meerakker}}]{Onvlee:CPC17:3583}
\bibinfo{author}{\bibfnamefont{J.}~\bibnamefont{Onvlee}},
  \bibinfo{author}{\bibfnamefont{S.~N.} \bibnamefont{Vogels}},
  \bibnamefont{and} \bibinfo{author}{\bibfnamefont{S.~Y.~T.}
  \bibnamefont{van~de Meerakker}}, \bibinfo{journal}{ChemPhysChem}
  \textbf{\bibinfo{volume}{17}}, \bibinfo{pages}{3583} (\bibinfo{year}{2016}).

\bibitem[{\citenamefont{Vogels et~al.}(2015)\citenamefont{Vogels, Onvlee,
  Chefdeville, van~der Avoird, Groenenboom, and van~de
  Meerakker}}]{Vogels:SCIENCE350:787}
\bibinfo{author}{\bibfnamefont{S.~N.} \bibnamefont{Vogels}},
  \bibinfo{author}{\bibfnamefont{J.}~\bibnamefont{Onvlee}},
  \bibinfo{author}{\bibfnamefont{S.}~\bibnamefont{Chefdeville}},
  \bibinfo{author}{\bibfnamefont{A.}~\bibnamefont{van~der Avoird}},
  \bibinfo{author}{\bibfnamefont{G.~C.} \bibnamefont{Groenenboom}},
  \bibnamefont{and} \bibinfo{author}{\bibfnamefont{S.~Y.~T.}
  \bibnamefont{van~de Meerakker}}, \bibinfo{journal}{Science}
  \textbf{\bibinfo{volume}{350}}, \bibinfo{pages}{787} (\bibinfo{year}{2015}).

\bibitem[{\citenamefont{Vogels et~al.}(2018)\citenamefont{Vogels, Karman,
  K{\l}os, Besemer, Onvlee, van~der Avoird, Groenenboom, and van~de
  Meerakker}}]{Vogels:NatureChem10:435}
\bibinfo{author}{\bibfnamefont{S.~N.} \bibnamefont{Vogels}},
  \bibinfo{author}{\bibfnamefont{T.}~\bibnamefont{Karman}},
  \bibinfo{author}{\bibfnamefont{J.}~\bibnamefont{K{\l}os}},
  \bibinfo{author}{\bibfnamefont{M.}~\bibnamefont{Besemer}},
  \bibinfo{author}{\bibfnamefont{J.}~\bibnamefont{Onvlee}},
  \bibinfo{author}{\bibfnamefont{A.}~\bibnamefont{van~der Avoird}},
  \bibinfo{author}{\bibfnamefont{G.~C.} \bibnamefont{Groenenboom}},
  \bibnamefont{and} \bibinfo{author}{\bibfnamefont{S.~Y.~T.}
  \bibnamefont{van~de Meerakker}}, \bibinfo{journal}{Nature Chemistry}
  \textbf{\bibinfo{volume}{10}}, \bibinfo{pages}{435} (\bibinfo{year}{2018}).

\bibitem[{\citenamefont{Gao et~al.}(2018)\citenamefont{Gao, Karman, Vogels,
  Besemer, van~der Avoird, Groenenboom, and van~de
  Meerakker}}]{Gao:NatureChem10:469}
\bibinfo{author}{\bibfnamefont{Z.}~\bibnamefont{Gao}},
  \bibinfo{author}{\bibfnamefont{T.}~\bibnamefont{Karman}},
  \bibinfo{author}{\bibfnamefont{S.~N.} \bibnamefont{Vogels}},
  \bibinfo{author}{\bibfnamefont{M.}~\bibnamefont{Besemer}},
  \bibinfo{author}{\bibfnamefont{A.}~\bibnamefont{van~der Avoird}},
  \bibinfo{author}{\bibfnamefont{G.~C.} \bibnamefont{Groenenboom}},
  \bibnamefont{and} \bibinfo{author}{\bibfnamefont{S.~Y.~T.}
  \bibnamefont{van~de Meerakker}}, \bibinfo{journal}{Nature Chemistry}
  \textbf{\bibinfo{volume}{10}}, \bibinfo{pages}{469} (\bibinfo{year}{2018}).

\bibitem[{\citenamefont{Akerman et~al.}(2017)\citenamefont{Akerman, Karpov,
  Segev, Bibelnik, Narevicius, and Narevicius}}]{Akerman:PRL119.073204}
\bibinfo{author}{\bibfnamefont{N.}~\bibnamefont{Akerman}},
  \bibinfo{author}{\bibfnamefont{M.}~\bibnamefont{Karpov}},
  \bibinfo{author}{\bibfnamefont{Y.}~\bibnamefont{Segev}},
  \bibinfo{author}{\bibfnamefont{N.}~\bibnamefont{Bibelnik}},
  \bibinfo{author}{\bibfnamefont{J.}~\bibnamefont{Narevicius}},
  \bibnamefont{and}
  \bibinfo{author}{\bibfnamefont{E.}~\bibnamefont{Narevicius}},
  \bibinfo{journal}{Phys. Rev. Lett.} \textbf{\bibinfo{volume}{119}},
  \bibinfo{pages}{073204} (\bibinfo{year}{2017}).

\bibitem[{\citenamefont{Liu et~al.}(2017)\citenamefont{Liu, Vashishta,
  Djuricanin, Zhou, Zhong, Mittertreiner, Carty, and
  Momose}}]{Liu:PRL118:093201}
\bibinfo{author}{\bibfnamefont{Y.}~\bibnamefont{Liu}},
  \bibinfo{author}{\bibfnamefont{M.}~\bibnamefont{Vashishta}},
  \bibinfo{author}{\bibfnamefont{P.}~\bibnamefont{Djuricanin}},
  \bibinfo{author}{\bibfnamefont{S.}~\bibnamefont{Zhou}},
  \bibinfo{author}{\bibfnamefont{W.}~\bibnamefont{Zhong}},
  \bibinfo{author}{\bibfnamefont{T.}~\bibnamefont{Mittertreiner}},
  \bibinfo{author}{\bibfnamefont{D.}~\bibnamefont{Carty}}, \bibnamefont{and}
  \bibinfo{author}{\bibfnamefont{T.}~\bibnamefont{Momose}},
  \bibinfo{journal}{Phys. Rev. Lett.} \textbf{\bibinfo{volume}{118}},
  \bibinfo{pages}{093201} (\bibinfo{year}{2017}).

\bibitem[{\citenamefont{Jansen et~al.}(2016)\citenamefont{Jansen, Semeria, and
  Merkt}}]{Jansen:JMolSpec322:9}
\bibinfo{author}{\bibfnamefont{P.}~\bibnamefont{Jansen}},
  \bibinfo{author}{\bibfnamefont{L.}~\bibnamefont{Semeria}}, \bibnamefont{and}
  \bibinfo{author}{\bibfnamefont{F.}~\bibnamefont{Merkt}}, \bibinfo{journal}{J.
  Mol. Spec.} \textbf{\bibinfo{volume}{332}}, \bibinfo{pages}{9}
  (\bibinfo{year}{2016}).

\bibitem[{\citenamefont{van~der Poel et~al.}(2015)\citenamefont{van~der Poel,
  Dulitz, Softley, and Bethlem}}]{Poel:NJP17:055012}
\bibinfo{author}{\bibfnamefont{A.~P.~P.} \bibnamefont{van~der Poel}},
  \bibinfo{author}{\bibfnamefont{K.}~\bibnamefont{Dulitz}},
  \bibinfo{author}{\bibfnamefont{T.~P.} \bibnamefont{Softley}},
  \bibnamefont{and} \bibinfo{author}{\bibfnamefont{H.~L.}
  \bibnamefont{Bethlem}}, \bibinfo{journal}{New J. Phys.}
  \textbf{\bibinfo{volume}{17}}, \bibinfo{pages}{055012}
  (\bibinfo{year}{2015}).

\bibitem[{\citenamefont{Cremers et~al.}(2018)\citenamefont{Cremers,
  Chefdeville, Plomp, Janssen, Sweers, and van~de
  Meerakker}}]{Cremers:PRA98:033406}
\bibinfo{author}{\bibfnamefont{T.}~\bibnamefont{Cremers}},
  \bibinfo{author}{\bibfnamefont{S.}~\bibnamefont{Chefdeville}},
  \bibinfo{author}{\bibfnamefont{V.}~\bibnamefont{Plomp}},
  \bibinfo{author}{\bibfnamefont{N.}~\bibnamefont{Janssen}},
  \bibinfo{author}{\bibfnamefont{E.}~\bibnamefont{Sweers}}, \bibnamefont{and}
  \bibinfo{author}{\bibfnamefont{S.~Y.~T.} \bibnamefont{van~de Meerakker}},
  \bibinfo{journal}{Phys. Rev. A} \textbf{\bibinfo{volume}{98}},
  \bibinfo{pages}{033406} (\bibinfo{year}{2018}).

\bibitem[{\citenamefont{Cremers et~al.}(2017)\citenamefont{Cremers,
  Chefdeville, Janssen, Sweers, Koot, Claus, and van~de
  Meerakker}}]{Cremers:PRA95:043415}
\bibinfo{author}{\bibfnamefont{T.}~\bibnamefont{Cremers}},
  \bibinfo{author}{\bibfnamefont{S.}~\bibnamefont{Chefdeville}},
  \bibinfo{author}{\bibfnamefont{N.}~\bibnamefont{Janssen}},
  \bibinfo{author}{\bibfnamefont{E.}~\bibnamefont{Sweers}},
  \bibinfo{author}{\bibfnamefont{S.}~\bibnamefont{Koot}},
  \bibinfo{author}{\bibfnamefont{P.}~\bibnamefont{Claus}}, \bibnamefont{and}
  \bibinfo{author}{\bibfnamefont{S.~Y.~T.} \bibnamefont{van~de Meerakker}},
  \bibinfo{journal}{Phys. Rev. A} \textbf{\bibinfo{volume}{95}},
  \bibinfo{pages}{043415} (\bibinfo{year}{2017}).

\bibitem[{\citenamefont{Wiederkehr et~al.}(2010)\citenamefont{Wiederkehr,
  Hogan, and Merkt}}]{Wiederkehr:PRA82:043428}
\bibinfo{author}{\bibfnamefont{A.~W.} \bibnamefont{Wiederkehr}},
  \bibinfo{author}{\bibfnamefont{S.~D.} \bibnamefont{Hogan}}, \bibnamefont{and}
  \bibinfo{author}{\bibfnamefont{F.}~\bibnamefont{Merkt}},
  \bibinfo{journal}{Phys. Rev. A} \textbf{\bibinfo{volume}{82}},
  \bibinfo{pages}{043428} (\bibinfo{year}{2010}).

\bibitem[{\citenamefont{Dulitz et~al.}(2015)\citenamefont{Dulitz, Vanhaecke,
  and Softley}}]{Dulitz:PRA91:013409}
\bibinfo{author}{\bibfnamefont{K.}~\bibnamefont{Dulitz}},
  \bibinfo{author}{\bibfnamefont{N.}~\bibnamefont{Vanhaecke}},
  \bibnamefont{and} \bibinfo{author}{\bibfnamefont{T.~P.}
  \bibnamefont{Softley}}, \bibinfo{journal}{Phys. Rev. A}
  \textbf{\bibinfo{volume}{91}}, \bibinfo{pages}{013409}
  (\bibinfo{year}{2015}).

\bibitem[{\citenamefont{{O. Chubar and P. Elleaume and J.
  Chavanne}}(2009)}]{RADIA_4.29}
\bibinfo{author}{\bibnamefont{{O. Chubar and P. Elleaume and J. Chavanne}}},
  \emph{\bibinfo{title}{Radia 4.29}} (\bibinfo{year}{2009}).

\bibitem[{\citenamefont{Dunning and Hulet}(1996)}]{Dunning:EMPS29B}
\bibinfo{author}{\bibfnamefont{F.~B.} \bibnamefont{Dunning}} \bibnamefont{and}
  \bibinfo{author}{\bibfnamefont{R.~G.} \bibnamefont{Hulet}},
  \emph{\bibinfo{title}{Atomic, Molecular and Optical Physics: Atoms and
  Molecules}}, vol. \bibinfo{volume}{29B} (\bibinfo{publisher}{Academemic
  Press}, \bibinfo{year}{1996}).

\bibitem[{\citenamefont{Yan et~al.}(2013)\citenamefont{Yan, Claus, van
  Oorschot, Gerritsen, Eppink, van~de Meerakker, and
  Parker}}]{Yan:RSI84:023102}
\bibinfo{author}{\bibfnamefont{B.}~\bibnamefont{Yan}},
  \bibinfo{author}{\bibfnamefont{P.~F.~H.} \bibnamefont{Claus}},
  \bibinfo{author}{\bibfnamefont{B.~G.~M.} \bibnamefont{van Oorschot}},
  \bibinfo{author}{\bibfnamefont{L.}~\bibnamefont{Gerritsen}},
  \bibinfo{author}{\bibfnamefont{A.~T. J.~B.} \bibnamefont{Eppink}},
  \bibinfo{author}{\bibfnamefont{S.~Y.~T.} \bibnamefont{van~de Meerakker}},
  \bibnamefont{and} \bibinfo{author}{\bibfnamefont{D.~H.}
  \bibnamefont{Parker}}, \bibinfo{journal}{Rev. Sci. Instrum.}
  \textbf{\bibinfo{volume}{84}}, \bibinfo{eid}{023102} (\bibinfo{year}{2013}).

\bibitem[{\citenamefont{Ploenes et~al.}(2016)\citenamefont{Ploenes, Haas,
  Zhang, van~de Meerakker, and Willitsch}}]{Ploenes:RSI87:053305}
\bibinfo{author}{\bibfnamefont{L.}~\bibnamefont{Ploenes}},
  \bibinfo{author}{\bibfnamefont{D.}~\bibnamefont{Haas}},
  \bibinfo{author}{\bibfnamefont{D.}~\bibnamefont{Zhang}},
  \bibinfo{author}{\bibfnamefont{S.~Y.~T.} \bibnamefont{van~de Meerakker}},
  \bibnamefont{and}
  \bibinfo{author}{\bibfnamefont{S.}~\bibnamefont{Willitsch}},
  \bibinfo{journal}{Rev. Sci. Instrum.} \textbf{\bibinfo{volume}{87}},
  \bibinfo{pages}{053305} (\bibinfo{year}{2016}).

\end{thebibliography}

\end{document}